\documentclass[12pt,english]{article}
\usepackage[T1]{fontenc}
\usepackage[latin9]{inputenc}
\usepackage{geometry}
\usepackage{color}
\usepackage{array}
\usepackage{textcomp}
\usepackage{multirow}
\usepackage{amsmath}
\usepackage{amssymb}
\usepackage{graphicx}
\usepackage{authblk}

\makeatletter

\providecommand{\tabularnewline}{\\}

\makeatother

\usepackage{babel}
\begin{document}
\title{Performance of energy harvesters with parameter mismatch }
\date{}
\author[1]{Tomasz Burzy{\`n}ski}
\author[2]{Piotr Brzeski}
\author[3]{Przemys{\l}aw Perlikowski}
\affil[1,2,3]{Division of Dynamics, Lodz University of Technology}

\maketitle
\begin{abstract}
\noindent This study explores the impact of parameter mismatch on the stability
of cross-well motion in energy harvesters, using a basin stability
metric . Energy harvesters, essential for converting ambient energy
into electricity, increasingly incorporate multi-well systems to enhance
efficiency. However, these systems are sensitive to initial conditions
and parameter variations, which can affect their ability to sustain
optimal cross-well motion---a state associated with maximum power
output. Our analysis compared four harvester types under varying levels
of parameter mismatch, assessing resilience of the devices to parameter
variations. By identifying safe operating ranges within the excitation
parameter space, this study provides practical guidance for designing
robust, stable harvesters capable of maintaining cross-well motion
despite parameter uncertainties. These insights contribute to advancing
the reliability of energy harvesting devices in real-world applications
where parameter mismatches are inevitable.
\end{abstract}

\newpage

\section{Introduction}

Energy harvesting technology has witnessed rapid growth in popularity
over the past decades. SCOPUS data indicates an almost exponential
rise in interest since 2003. Specifically, the keyword \textquotedbl Energy
harvesting\textquotedbl{} appeared in approximately 250 papers in
2003, around 1300 papers in 2009, about 5000 papers in 2015, and close
to 9000 papers in 2023. The number of articles published up to July
2024 (around 9000) suggests that even more papers may be published
by the end of the year.

Energy harvesters are employed in various scientific fields to replace
conventional energy sources and generate electricity from ambient
vibrations. The primary objective of energy harvesting technology
is to maximize power generation under given excitation amplitudes
and frequencies. The frequency spectra considered vary based on the
intended use. For instance, energy harvesters for wearable devices
aim to operate in the low-frequency range $(1-10\mathrm{Hz})$, \cite{Human_freq_1,Human_freq_2},
similar to those used for capturing energy from ocean waves \cite{Ocena_freq_1}.
Conversely, energy scavenging from vehicles, such as trains, occurs
in the frequency spectrum of $20-65\mathrm{Hz}$ \cite{Train_freq_2,Train_freq_1}.
When the excitation frequency matches the device's natural frequency,
increased power output is observed which is manifested by spikes on
the frequency response curves (FRC). To achieve a more complex FRC,
with more spikes and higher power output, systems with multiple potential
wells are employed. Multi-well systems have became a standard in energy
harvesting technology, and recently harvesters are have become increasingly
sophisticated and non-linear to further extend the operation range.

Recent studies provide illustrative examples. Li et al. in \cite{Li2023}introduced
a classical bistable mechanical electromagnetic energy harvester augmented
with a non-linear boundary, incorporating strong Duffing-type non-linearities
to enhance energy harvesting. Zhang et all. \cite{Kolka_impactcs_assymetry}
tried to increase device performance by introducing a discontinuity
in the form of impacts to the model, the same applies to the work
of Wang et all. \cite{Wang2024}. Additional mass and springs are
added to the classical piezoelectric energy harvester by Chen et all
\cite{dual_coipling_warnenski}. A strong negative stiffness is introduced
to the classical harvester model by Chen and Zhao \cite{Zero_stif}
to create so-called quasi-zero stiffness oscillator and facilitate
energy scavenging. These examples demonstrate that various non-linearities
and discontinuities are introduced to enhance device performance,
albeit at the cost of increased multistability, which has become an
inherent aspect of energy harvesting from electromechanical devices.

Considering a bistable system with two potential wells, commonly used
in energy harvesting, such systems possess three equilibrium positions:
one unstable and two stable. These systems can exhibit qualitatively
different behaviors, including oscillations around one equilibrium
position, transitions between two equilibrium positions, oscillations
with significant amplitude encompassing both extreme equilibrium positions,
and chaotic motion. Cross-well motion between two extreme equilibrium
positions is most favorable for maximizing power output. This fact
was observed by Sosna et al. \cite{Train_freq_1} in piezoelectric
energy harvester. The same was reported by Li et al. \cite{corss_well_3}.
The electromagnetic device was analyzed with the same conclusions
by Chen et al. \cite{corss-Well_2}, here authors also emphasize
the drop in power output during chaotic motion. Tang et al. \cite{High_response_activation}
in his work introduced variable damping control to a piezoelectric
harvester to activate high-energy response that corresponds to cross-well
motion. Yan et al. \cite{corss_well_4_quad} studied a quad-stable
energy harvester with four stable equilibrium positions, finding optimal
operation during quad-well motion spanning extreme equilibrium positions.

Taking into consideration the examples given, the assessment of device's
effectiveness can be reduced to the ability to reach and maintain
cross-well motion. Because considered systems are multistable, hence
initial conditions sensitive, we focus on the probability of reaching
cross-well motion. When it comes to probabilistic analysis of multistable
systems, basin stability metrics, and sample-based approach are good
candidates for the measures used to assess the systems \cite{Leszczynski2022,Brzeski2017}.
The analysis may be further extended by adding parameters mismatch
\cite{BS_with_mismatch}, since harvesters include complex interactions
between mechanical and electrical subsystems that are challenging
to model.

The novelty in this study is to use basin stability type metric with
parameter mismatch to assess the effectiveness of the energy harvesters
based on the system response dynamics rather than on the power output.
The goal of the study is to answer the following questions: What are
the excitation parameters when the cross-well motion has the greatest
basin stability regardless of the parameters mismatch? How different
types of systems are prone to parameter mismatch? What degree of parameter
mismatch is significant? 

\section{Materials and methods}

In this paper, we focus on calculating the probability of reaching
stable cross-well motion in four different energy harvesters. For
this purpose, we employ the basin stability method with parameter
mismatch, first introduced by Brzeski et al. in \cite{BS_with_mismatch}
to detect and classify coexisting solutions in non-linear systems.
It is an extension of the original basin stability method introduced
by Menck et al. \cite{Menck2013}. Basin stability with parameter
mismatch assumes that the exact values of chosen parameters are unknown
due to finite precision, indirect measurement methods or their variability
during motion (e.g., dependence on displacement, velocity or ect.).
This results in an additional variance between the real-world system
and the numerical model, affecting the overall accuracy of the mathematical
approximation.

We examine how this variance influences the model by determining whether
the solution obtained for fixed parameter values remains the sole
possible attractor or at least maintains a major basin of attraction
when the chosen parameters fluctuate. This approach can be considered
a type of stability analysis, where we investigate how the system
dynamics may change due to system degradation or significant misjudgment
of parameter values.

In our study, we assume that two parameters (excitation amplitude
and frequency) define a 2D space where we investigate the system dynamics.
These parameters are drawn from predefined sets using the Monte Carlo
approach. Additionally, we vary other parameters that are difficult
to identify or prone to change as the system wears. For each parameter,
we assume a reference value, believed to be the real one, and observe
how the system dynamics change as these parameters vary by $\pm2.5\%$,
$\pm5\%$, and $\pm10\%$. For every system, we specify which parameters
are mismatched, but for most systems these are: viscous damping, electromechanical
coupling coefficients, and parameters defining the system's potential
energy. We choose these parameters because mechanical damping in all
considered systems is expressed by viscous damping. Energy dissipation
is a complex phenomenon, often involving multiple sophisticated processes
simultaneously alongside viscous damping. Therefore, we assume that
describing the entire phenomenon with a viscous model is challenging,
making damping coefficients good candidates for parameter mismatch.
The same applies to electromechanical coupling coefficients, which
are linearized and proportional to velocity/current, while the actual
interactions between mechanical and electrical subsystems are far
more complex (see, for example, \cite{Modelowanie_cewki_1}, where
the authors optimize the parameters of a mathematical model of a coil,
or \cite{Modelowanie_cewki_2}, where the electromagnetic coupling
coefficient is a function of the magnet's position relative to the
coil). Additionally, system stiffness may degrade as the system wears,
so we also consider stiffness parameters during mismatch.

\begin{figure}
\begin{centering}
\includegraphics{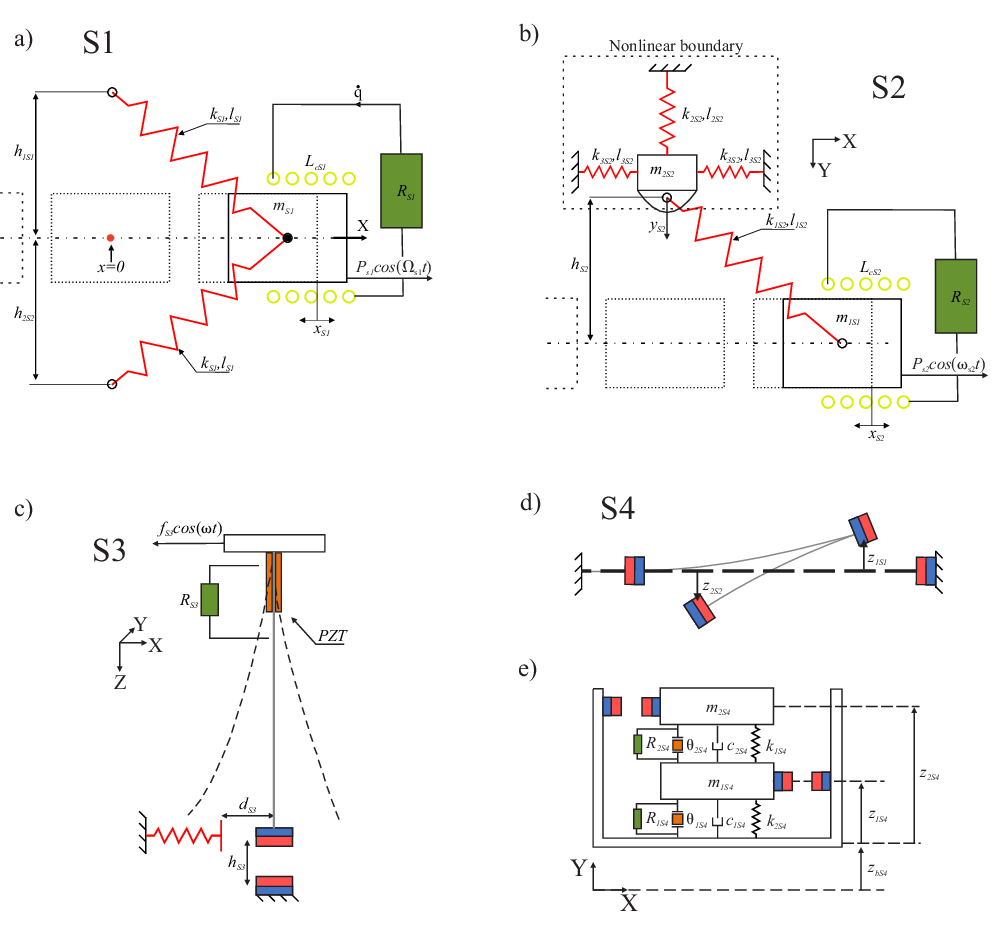}
\par\end{centering}
\caption{\label{fig:Physcial_systems}Schematic diagrams of four investigated
systems. a) Classical electromagnetic bistable energy harvester (S1).
b) Electromagnetic bistable energy harvester with non-linear boundary
(S2), introduced by \cite{Li2023}. c) Asymmetric piezoelectric energy
harvester (S3), introduced by \cite{Wang2024}. d,e) Two degrees
of freedom piezoelectric energy harvester (S4), introduced by \cite{Costa2024}.}

\end{figure}

In this paper, we investigate four energy harvesters, two of electromagnetic
and two of piezoelectric type. The first model is derived from the
beginning, while the others are introduced and derived in referenced
papers \cite{Costa2024,Li2023,Wang2024}. Detailed information about
these models can be found in those sources. Figure \ref{fig:Physcial_systems}
presents schematic diagrams of all systems, providing a comprehensive
comparison and showing their variety. This figure should not be treated
as a detailed representation of the physical models used for deriving
the mathematical models, as those details are available in the original
papers (except for the first system). Relevant descriptions of the
systems and their mathematical models are presented in the following
subsections.

\subsection{Classical bistable energy harvester (S1) }

\begin{figure}
\begin{centering}
\includegraphics{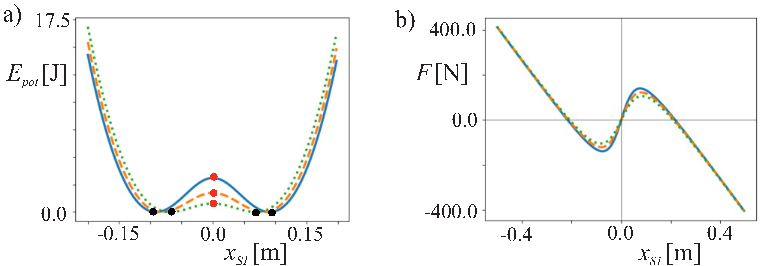}
\par\end{centering}
\caption{\label{fig:a)-Elastic-potential}a) Elastic potential energy of S1
for different values of $h$ parameter. Red dots represents unstable
equilibrium positions, black dots stable equilibrium positions. Blue
solid line is for $h_{1S1}=h_{2S1}=0.6l_{0}$ , orange dashed for
$h_{1S1}=h_{2S1}=0.7l_{0}$, green dotted for $h_{1S1}=h_{2S1}=0.8l_{0}$
b) Restoring force (in $X$ direction) generated by spring along $X$
axis.}

\end{figure}

The first system (S1) is a bistable mechanical oscillator. The physical
model of the device is presented in \ref{fig:Physcial_systems}a.
Bistability results from the geometry of the system that consists
only of elements with linear response. Mass attached to two springs
oscillates and moves the magnet inside a coil. The stiffness of the
system is controlled by varying the distance between the spring mounting
point and the translational axis of the oscillator - parameter $h$.
By proper parameter tuning, we can obtain strong non-linear stiffness.
System configuration presented in \ref{fig:Physcial_systems}(a) demonstrates
three equilibrium positions marked with dots on the X axis. A red
dot means unstable equilibrium, black dot stable equilibrium (only
one stable equilibrium is presented but the other one, symmetric,
exists on the other side). Figure \ref{fig:a)-Elastic-potential}
shows the elastic potential energy of the device and the stiffness
of the system along the $\mathrm{X}$ axis for different values of
$h$ parameter. We can observe a typical double-well potential landscape
and the phenomena of negative stiffness that is typical for such a
system. In this system configuration, we may observe different behaviors
like oscillations around one of the equilibrium positions, jumping
between two different equilibrium positions, and oscillations with
significant amplitude spanning both extreme equilibrium positions.

The considered system consists of electrical and mechanical subsystems.
The mathematical model possesses two generalized coordinates - mechanical
variable $x$ which is the translational displacement of the mass
attached to the springs and electrical variable $q$ which is an electric
charge. The magnet moving inside the coil is a coupling terminal that
demonstrates the interaction between the mechanical and electrical
subsystems. The coupling is originated from the magnetic energy of
the system. For the electrical subsystem, we define magnetic coenergy,
derived according to \cite{Lagrangian}, as:

\begin{equation}
W=\frac{L_{cS1}\dot{q}^{2}}{2}+\alpha_{S1}\dot{q}x,\label{eq:1}
\end{equation}

\noindent where, $L_{c}$ is inductance of the coil, $\alpha$ is
electromagnetic force constant. Translational kinetic energy and elastic
potential energy for the system are: 

\[
T=\frac{m_{S1}\ddot{x}}{2},
\]

\[
V=k_{S1}\left(l_{S1}-\sqrt{x^{2}+h_{1S1}^{2}}\right)^{2}+k_{S1}\left(l_{S1}-\sqrt{x^{2}+h_{2S2}^{2}}\right)^{2}.
\]

\noindent Generalized forces for the mechanical and electrical subsystems
are presented below: 

\[
Q_{m}=A\cos(\omega t)-d\dot{x},
\]

\[
Q_{e}=-R_{S1}\dot{q},
\]

\noindent where, $A$ is an amplitude of the external mechanical excitation,
$d$ is the coefficient of energy dissipation due to internal resistances,
$R_{S1}$ is the sum of electrical resistance in the system. Equations
of motion are derived using Lagrange's equations of second kind: 

\begin{equation}
\frac{d}{dt}\left(\frac{\partial\mathcal{L}}{\partial\dot{x}}\right)-\frac{\partial\mathcal{L}}{\partial x}=Q_{m},\label{eq:Lag_mech}
\end{equation}

\begin{equation}
\frac{d}{dt}\left(\frac{\partial\mathcal{L}}{\partial\dot{q}}\right)-\frac{\partial\mathcal{L}}{\partial q}=Q_{e},\label{eq:Lag_ele}
\end{equation}

\noindent where Lagrangian is defined as $\mathcal{L}=T-V+W$. Using
following substitutions for the general coordinates: $\tau=\omega_{nS1}t$,
$x=\bar{x}l_{0}$, $\dot{q}=\dot{\bar{q}}i_{0}$, where, $\omega_{nS1}=\sqrt{\frac{k_{S1}}{m_{S1}}}$,
$l_{S1}$ is the free length of the spring, and $i_{0}$ is some reference
current, equations are made dimensionless ($\dot{}$ means derivative
with respect to $\tau$): 

\[
\ddot{\bar{x}}+\bar{x}\left(1-\frac{1}{\sqrt{\bar{x}^{2}+\gamma_{1}^{2}}}\right)+\bar{x}\left(1-\frac{1}{\sqrt{\bar{x}^{2}+\gamma_{2}^{2}}}\right)-\theta\dot{q}=P_{S1}\cos(\varOmega_{S1}\tau)-\varphi\dot{\bar{x}},
\]

\[
\ddot{\bar{q}}+\varepsilon\dot{\bar{x}}+\lambda\dot{q}=0,
\]

\noindent including dimensionless parameters:

\[
\varOmega=\frac{\omega}{\omega_{nS1}},\quad\gamma_{1}=\frac{h_{1}}{l_{0}},\quad\gamma_{2}=\frac{h_{2}}{l_{0}},\quad\theta=\frac{\alpha_{S1}i_{0}}{l_{0}k_{1S1}},\quad\varphi_{S1}=\frac{D}{\sqrt{k_{S1}m_{S1}}},\quad\varepsilon=\frac{\alpha_{S1}l_{0}}{L_{cS1}i_{0}},
\]

\[
\lambda_{S1}=\frac{R_{S1}}{L_{cS1}\sqrt{\frac{k_{S1}}{m_{S1}}}},\quad P=\frac{A}{l_{0}k_{S1}}
\]

\noindent where: 

\[
m_{S1}=0.2\;\left[kg\right],\quad l_{0}=0.114\;\left[m\right],h_{1}=h_{2}=0.95l_{0}\;\left[m\right],L_{cS1}=1.463\;\left[H\right],
\]

\[
\alpha_{S1}=30\;\left[\frac{N}{A}\right],\quad c_{S1}=0.35\;\left[\frac{Ns}{m}\right],\quad k_{S1}=1500\;\left[\frac{N}{m}\right],\quad R_{S1}=2200\;\left[\Omega\right].
\]

\noindent Values of electrical subsystem parameters $\alpha_{S1}$,
$L_{cS1}$, $R_{S1}$ were taken from \cite{Kecik2017} for a similar
real-world electromagnetic system. Spring parameters $k_{S1}$, $l_{0}$
are obtained by measuring existing products whose size and stiffness
correspond to the system dimensions. The chosen value of $m_{S1}$
is not overloading spring elements and still corresponds in size to
the system presented in \cite{Kecik2017}. Mechanical damping component
$c_{S1}$ is set to $2.5\%$ of critical damping of the system since
extremely low friction joints need to be used to assure the efficiency
of such a device. 

\subsection{Bistable energy harvester with non-linear elastic boundary (S2)}

The second system (S2) is the expansion of the first one. System S2
also consists of a magnet inside a coil supported by the hinged spring
in such a way that strong geometric non-linearity is created. What
distinguishes the two systems is the non-linear elastic boundary.
In S2 the boundary is introduced by a set of additional springs ($k_{2S2}$,
$k_{3S2}$) that support the hinge. In S1 the hinged spring was assumed
to have infinite stiffness support, in S2 the support is elastic.
The system was introduced for the first time by Li et al. in \cite{Li2023},
and det\textcolor{black}{ailed model descriptions can be found there.
Governing dimensionless equations of motion are as follows:}

\[
\ddot{\mathrm{X}}+\mathrm{X}\left(1-\frac{1}{\sqrt{\mathrm{X}^{2}+\eta_{1}^{2}\left(1-\mathrm{Y}\right)^{2}}}\right)+\varphi_{1S2}\dot{\mathrm{X}}=P_{S2}\cos(\omega_{S2}t)-\rho\mathrm{I}
\]

\[
\ddot{\mathrm{Y}}+\frac{\lambda}{\mu_{1}}\mathrm{Y}+2\frac{\lambda}{\mu_{2}}\mathrm{Y}\left(1-\frac{1}{\sqrt{1+\eta_{2}^{2}\mathrm{Y}^{2}}}\right)-\lambda\left(1-\mathrm{Y}\right)\left(1-\frac{1}{\sqrt{\mathrm{X}^{2}+\eta_{1}^{2}\left(1-\mathrm{Y}\right)^{2}}}\right)+\varphi_{2S2}\dot{\mathrm{Y}}=0
\]

\[
\mathrm{\dot{I}}+\theta_{S2}\mathrm{I}=\varepsilon\mathrm{\dot{X}}
\]

\noindent where, $\varphi_{iS2}$ are dimensionless damping coefficients
of the the 1st and 2nd mechanical degree of freedom, $\eta_{1}=\frac{h_{S2}}{l_{1}}$,
$\eta_{2}=\frac{h_{S2}}{l_{3}}$ (see the figure \ref{fig:Physcial_systems}b),
$P$ is dimensionless excitation amplitude, $\lambda$ is mass ratio,
$\rho$ and $\varepsilon$ are electromechanical coupling coefficients
for electrical and mechanical part respectively, $\mu_{1}=\frac{k_{1}}{k_{2}}$,
$\mu_{2}=\frac{k_{1}}{k_{3}}$ (see the figure \ref{fig:Physcial_systems}b).
Values of the parameters are the following:

\[
\lambda=4\quad\eta_{1}=0.92\quad\eta_{2}=1.75\quad\varphi_{1S2}=0.005\quad\varphi_{2S2}=0.02\quad\mu_{1}=\mu_{2}=1
\]

\[
\theta=20\quad\varepsilon_{S2}=13.13\quad\rho=0.005
\]

\subsection{Asymmetric piezoelectric bistable energy harvester (S3)}

The third system (S3) is based on the simple form of the piezoelectric
energy harvester with two repulsive magnets and a beam with a piezoelectric
transducer attached. Additionally, there is a unilateral spring that
limits the motion of the beam and introduces to the system asymmetry
and soft impacts. The model was introduced by Wang et al. in \cite{Wang2024}
and detailed description can be found there. The system is governed
by the following dimensionless equations: 

\[
\ddot{y}+2\xi_{1S3}\dot{y}-y+\beta_{S3}y+\delta_{S3}y^{3}+g\left(y,\dot{y}\right)-\kappa_{S3}^{2}V=f_{S3}\cos(\omega_{S3}t)
\]

\[
\dot{V}+\alpha_{S3}V+\dot{y}=0
\]

\[
g\left(y,\dot{y}\right)=\begin{cases}
0, & y>-d_{S3}\\
2\xi_{2S3}\dot{y}+K_{S3}\left(y+d_{S3}\right), & y\leq-d_{S3}
\end{cases}
\]

\noindent where, $\xi_{1S3}$ and $\xi_{2S3}$ are dimensionless damping
ratios, $\beta_{S3}$ and $\delta_{S3}$ are dimensionless coefficients
characterizing non-linear restoring forces, $\kappa_{S3}$ is dimensionless
electromechanical coupling coefficient, $d_{S3}$ is the dimensionless
distance marked in the figure \ref{fig:Physcial_systems}c, $K_{S3}$
is dimensionless collision stiffness, $\omega_{S3}$ is dimensionless
excitation frequency, and $f_{S3}$ is dimensionless excitation amplitude.
The values of parameters are as follows:

\[
\xi_{1S3}=0.08\quad\xi_{2S3}=0.05\quad\beta_{S3}=0.25\quad\delta_{S3}=0.5\quad\kappa_{S3}=\sqrt{0.002}\quad\alpha_{S3}=0.4\quad K_{S3}=100\quad d_{S3}=0.6
\]

\subsection{Compact non-linear piezoelectric energy harvester (S4)}

Fourth system (S4) is a two-degree-of-freedom piezoelectric energy
harvester that consists of two coupled beams and four magnets as indicated
in the figure \ref{fig:Physcial_systems}d and e. It is an expansion
of the system S3 but without the asymmetric spring. This system is
proposed by Costa et al. in \cite{Costa2024} and detailed information
about the system can be found there. Dimensionless equations of motion
of the system are given below: 

\[
\ddot{\bar{z}}_{1}+2\zeta_{1S4}\dot{\bar{z}}_{1}-2\zeta_{2S4}\left(\dot{\bar{z}}_{2}-\dot{\bar{z}}_{1}\right)+\left(1+\alpha_{1S4}\right)\bar{z}_{1}+\beta_{1S4}\bar{z}_{1}-\rho_{S4}\varOmega_{s}^{2}\left(\bar{z}_{2}-\bar{z}_{1}\right)-\chi_{1S4}\bar{\upsilon}_{1}+\chi_{2S4}\bar{\upsilon}_{2}=-\ddot{\bar{z}}_{b}
\]

\[
\rho_{S4}\ddot{\bar{z}}_{2}+2\zeta_{2S4}\left(\dot{\bar{z}}_{2}-\dot{\bar{z}}_{1}\right)+\alpha_{2S4}\bar{z}_{2}+\beta_{2S4}\bar{z}_{2}+\rho_{S4}\varOmega_{s}\left(\bar{z}_{2}-\bar{z}_{1}\right)-\chi_{2S4}\bar{\upsilon}_{2}=-\ddot{\bar{z}}_{b}
\]

\[
\dot{\bar{\upsilon}}_{1}+\varphi_{1S4}\bar{\upsilon}_{1}+\kappa_{1S4}\dot{\bar{z}}_{1}=0
\]

\[
\dot{\bar{\upsilon}}_{2}+\varphi_{2S4}\bar{\upsilon}_{2}+\kappa_{2S4}\left(\dot{\bar{z}}_{2}-\dot{\bar{z}}_{1}\right)=0
\]

\noindent where, $z_{b}=\gamma_{S4}\sin(\varOmega_{S4}\tau)$ is the
dimensionless excitation frequency, $\gamma_{S4}$ is the dimensionless
excitation amplitude, $\varOmega_{S4}$ is the dimensionless excitation
frequency, $\rho_{S4}$ is ratio of masses, $\zeta_{iS4}$ are dimensionless
mechanical damping coefficients of 1st and 2nd mechanical degree of
freedom, $\varOmega_{s}$ is ratio of linearized natural frequencies,
$\alpha_{iS4}$ are dimensionless linear restitution coefficients,
$\beta_{iS4}$ are dimensionless non-linear restitution coefficients,
$\chi_{iS4}$ are dimensionless piezoelectric coupling coefficients
in the mechanical ODE, $\kappa_{iS4}$ are dimensionless piezoelectric
coupling coefficients in the electrical ODE, $\varphi_{iS4}$ are
dimensionless electrical resistance of the 1st and 2nd circuits. Values
of parameters are as follows: 

\[
\zeta_{1S4}=\zeta_{2S4}=0.025\quad\alpha_{1S4}=-2\quad\alpha_{2S4}=-1\quad\beta_{1S4}=\beta_{2S4}=1
\]

\[
\chi_{1S4}=\chi_{2S4}=0.05\quad\kappa_{1S4}=\kappa_{2S4}\quad\rho_{S4}=1\quad\varOmega_{s}=0.25\quad\varphi_{1S4}=\varphi_{2S4}=0.05
\]

\section{\label{sec:Results}Results}

This section presents basin stability with parameter mismatch metrics
for different systems. Their dynamics is investigated in 2D parameter
space that is divided into 1600 boxes. The color intensity of each
box determines the probability of reaching cross-well motion in that
part of the parameter space. The probability varies from 0 to 1, where
1 means that 100\% of all samples demonstrate cross-well motion, otherwise
there are other attractors. Each system is simulated four times (for
different amounts of parameter mismatch) and each simulation includes
200000 samples. The number of samples was selected in such a way that
each box contained at least 100 samples. Dimensionless excitation
frequency and amplitude are presented on the horizontal and vertical
axis respectively. Figure captions tell which other parameters were
mismatched and what was the range of initial conditions considered. 

\begin{figure}
\begin{centering}
\includegraphics{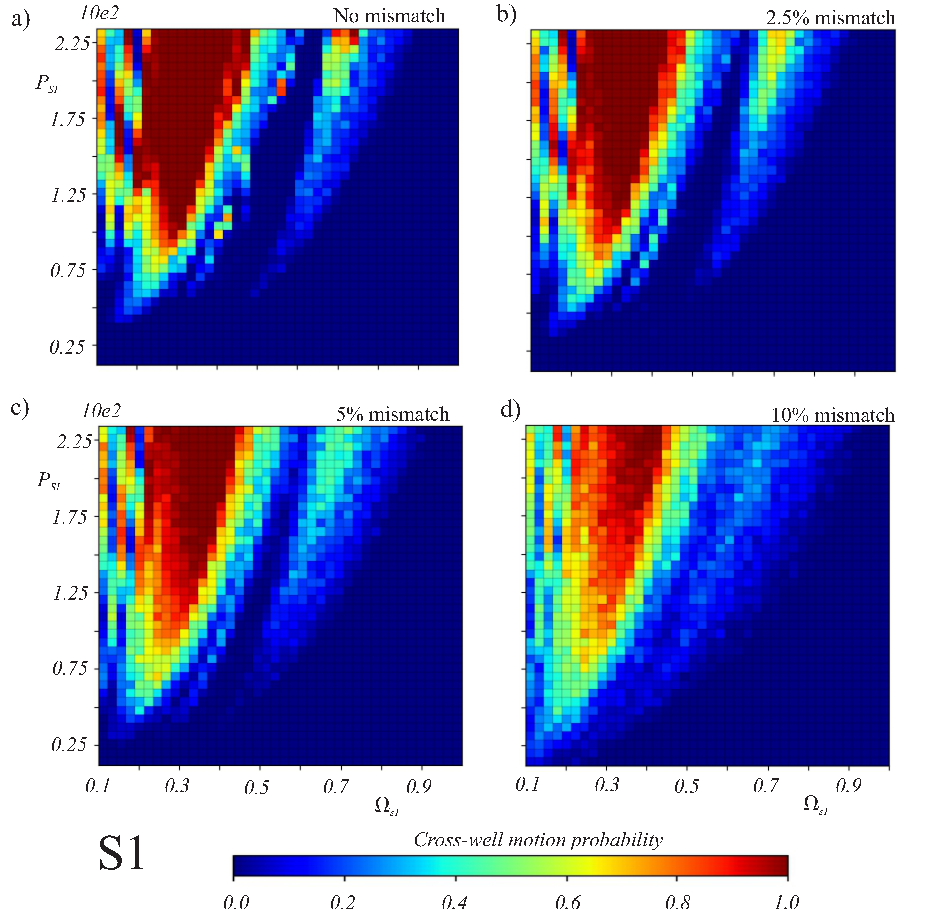}
\par\end{centering}
\caption{\label{fig:S1_res}Probability of reaching stable cross-well motion
for S1. Parameters that are mismatched: $\gamma_{2S1}$, $\theta_{S1}$,
$\varepsilon_{S1}$, $\varphi_{S1}$ represent respectively: potential
well shape, electromechanical coupling, and mechanical damping. Range
of initial conditions: $\overline{x}_{S1}$, $\dot{\overline{x}}_{S1}\epsilon<-1,1>$
Degree of parameter mismatch: a) No mismatch b) $\pm2.5\%$ c) $\pm5\%$
d) $\pm10\%$.}
\end{figure}

\begin{figure}
\begin{centering}
\includegraphics{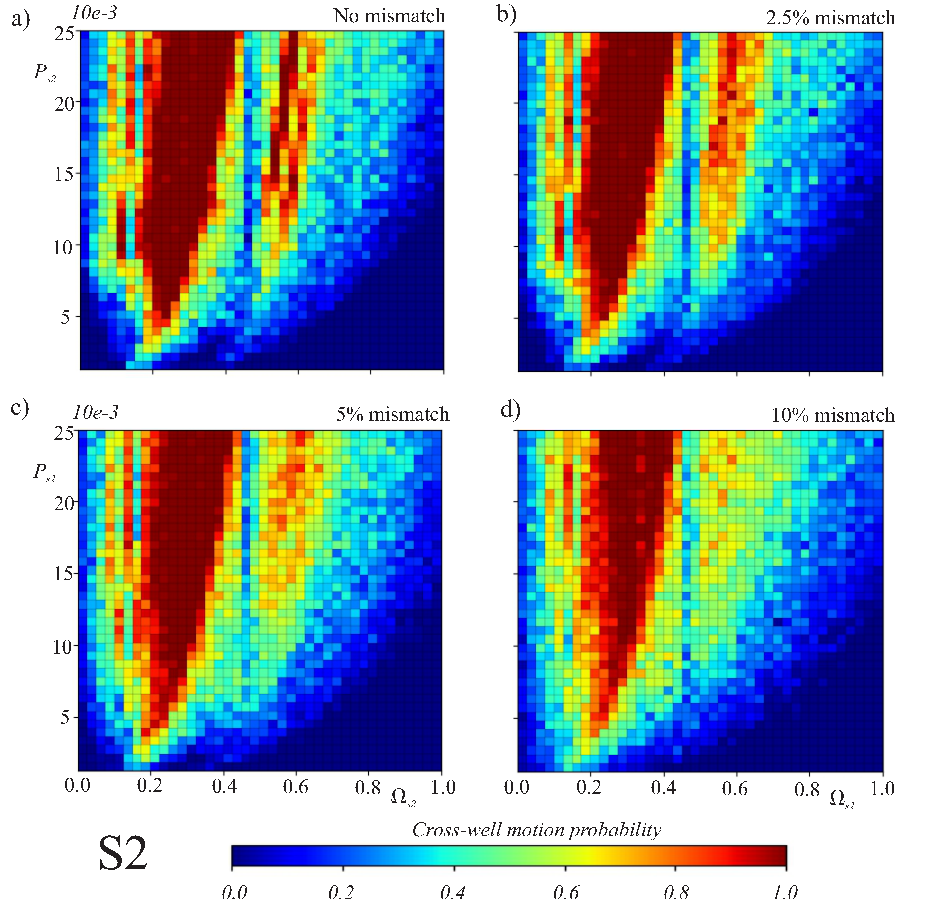}
\par\end{centering}
\caption{\label{fig:S2_res}Probability of reaching stable cross-well motion
for S2. Parameters that are mismatched:$\eta_{1S2}$, $\eta_{2S2}$,
$\rho_{S2}$, $\varepsilon_{S2}$, $\varphi_{1S2}$, $\varphi_{1S2}$
represent respectively: potential well shape, electromechanical coupling,
and mechanical damping. Range of initial conditions: $\mathrm{X}$,
$\dot{\mathrm{X}}\epsilon<-1,1>$ Degree of parameters mismatch: a)
No mismatch b) $\pm2.5\%$ c) $\pm5\%$ d) $\pm10\%$.}

\end{figure}

\begin{figure}
\begin{centering}
\includegraphics{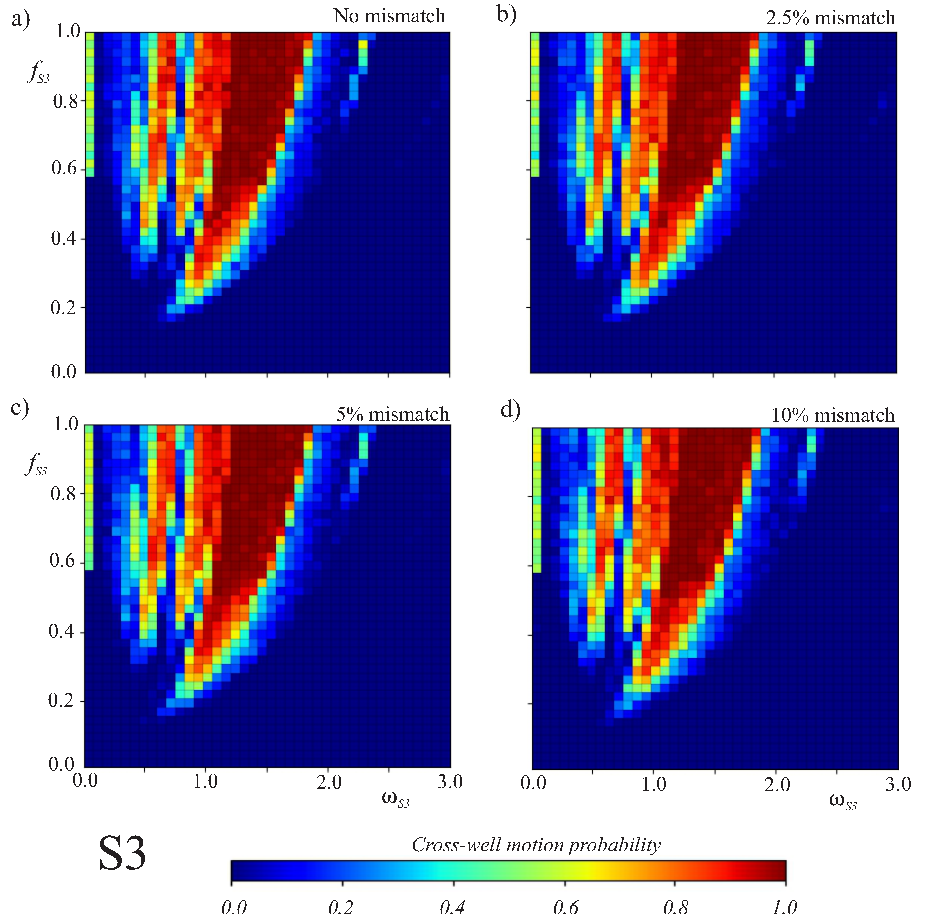}
\par\end{centering}
\caption{\label{fig:S3_res}Probability of reaching stable cross-well motion
for S3. Parameters that are mismatched:$K_{S3}$, $\kappa_{S3}$,
$\zeta_{1S3}$, $\zeta_{2S3}$ represent respectively: stiffness of
the unilateral stop, electromechanical coupling, and mechanical damping.
Range of initial conditions: $\mathrm{y}\epsilon<-0.55,2>$, $\dot{\mathrm{y}}\epsilon<-2,2>$
Degree of parameter mismatch: a) No mismatch b) $\pm2.5\%$ c) $\pm5\%$
d) $\pm10\%$.}

\end{figure}

\begin{figure}
\begin{centering}
\includegraphics{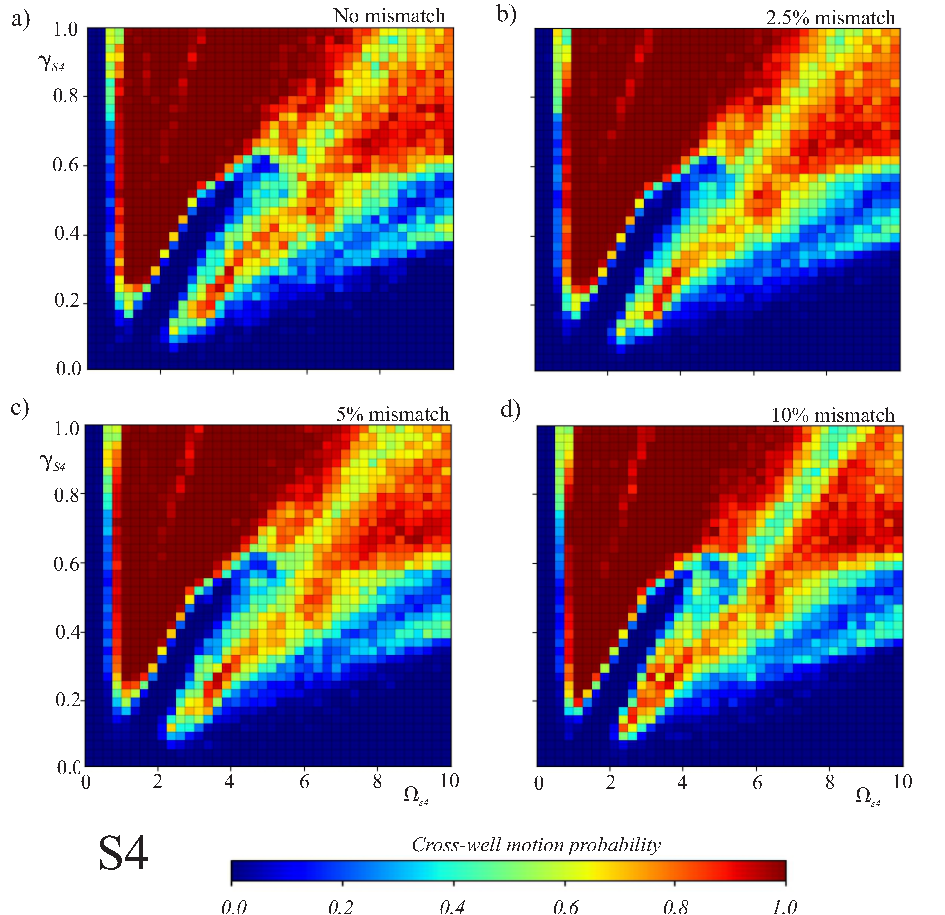}
\par\end{centering}
\caption{\label{fig:S4_res}Probability of reaching stable cross-well motion
for S4. Parameters that are mismatched:$\chi_{1S4}$, $\chi_{2S4}$,
$\kappa_{1S4}$, $\kappa_{2S4}$, $\zeta_{1S4}$, $\zeta_{2S4}$ represent
respectively: electromechanical coupling coefficients, and mechanical
damping. Range of initial conditions: $\mathrm{z_{1}}\epsilon<-2,2>$,
$z_{2}\epsilon<-2,2>$ Degree of parameter mismatch: a) No mismatch
b) $\pm2.5\%$ c) $\pm5\%$ d) $\pm10\%$.}

\end{figure}

\begin{figure}
\begin{centering}
\includegraphics{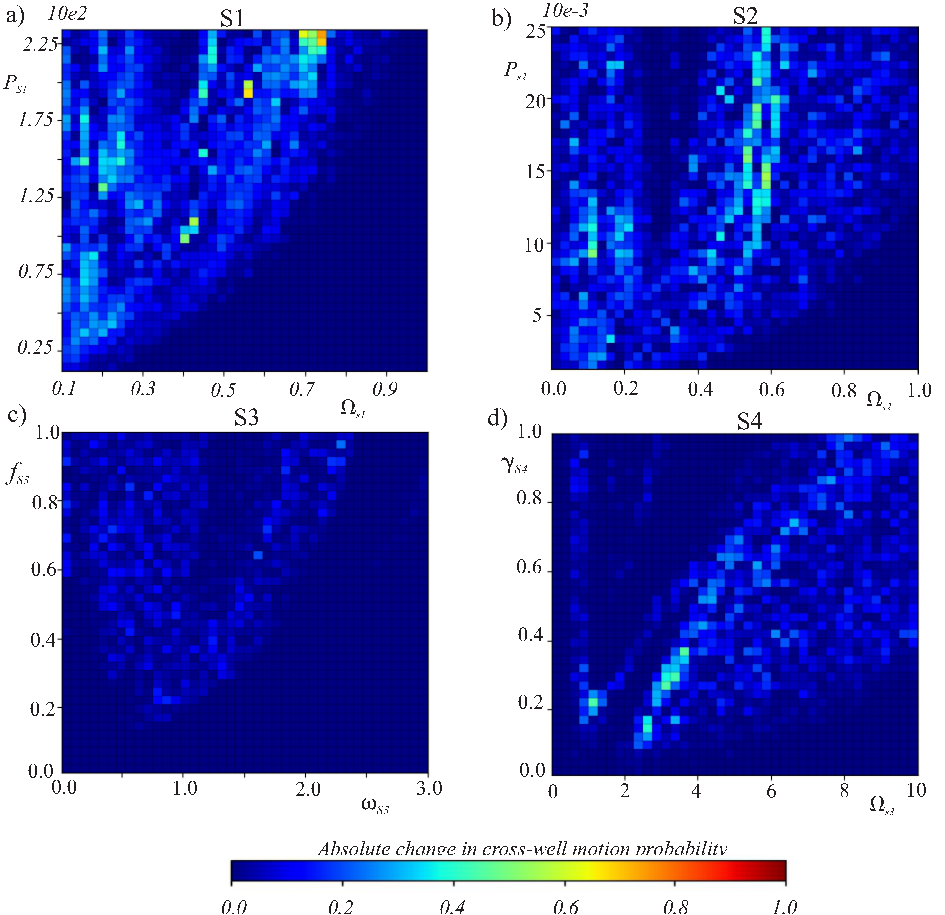}
\par\end{centering}
\caption{\label{fig:compare_prob}Change in cross-well motion probability while
the parameters were mismatched $\pm10\%$ a) S1 b) S2 c) S3 d) S4}

\end{figure}

Figure \ref{fig:S1_res} shows results for S1. Panel (a) represents
situation when all other parameters than excitation amplitude and
frequency are fixed. We can observe a tongue-shaped region spanned
from $\Omega_{S1}\epsilon<0.2,0.4>$ and $P_{S1}>1$ where the probability
of reaching cross-well motion is 1. If we want to maximize the efficiency
of the device it should operate in this parameter range. Panels (b),
(c), and (d) show situations when we vary other parameters. These
parameters are indicated in the figure caption. We observe the erosion
of the probability of reaching stable cross-well motion. The greater
the parameter mismatch, the greater the erosion. In panel (d) the
regions with a 100\% probability of cross-well motion are found only
when $\Omega_{S1}=0.4$ and $P_{S1}>1.75$. Erosion progresses with
increasing excitation amplitude and frequency values. In practice,
it means that if the system was excited with $\Omega_{S1}=0.3$ and
$P_{S1}=1.25$ assuming exact values of other parameters the model
assures that the stable cross-well motion is the only existing attractor.
However, if it turns up that the real values of chosen parameters
(indicated in figure caption) are within 10\% of assumed values the
probability drops to $\thickapprox70\%$. Also the second tongue-shaped
region with increased probability visible in figure \ref{fig:S1_res}(a)
almost completely disappears in panel (d) of the same figure, indicating
that it is susceptible to parameter change. 

Figure \ref{fig:S2_res} shows results for S2. Panel (a) represents
a situation when all other parameters except excitation amplitude
and frequency are fixed. The range of considered excitation frequency
and amplitude corresponds to the values analyzed in the original paper
\cite{Li2023}. The tongue-shaped region with the red color visible
in panel (a) indicates which parameters provide the existence of solely
the desired cross-well solution. S2 guarantees reaching this solution
in a greater parameter range compared to S1. The model is more complex
and stronger nonlinearity exists due to elastic boundary. However,
similarly as in the previous case, we observe the erosion of probability
of reaching cross-well motion as we increase parameter mismatch. Erosion
progresses with increasing excitation amplitude and frequency values.
The second smaller tongue does not disappear however the probability
drops. When the parameter mismatch reaches $\pm10\%$ the probability
decreases to $\thickapprox0.6$ while with fixed parameter values
there were regions with probability equal to $1$. The results for
S2 show that in practice the second smaller tongue should not be considered
an effective-operation region since it is sensitive to small parameter
variations. Figure \ref{fig:S2_res} shows what are the excitation
parameter values when the main tongue-shaped regions ensure only one
high-power attractor regardless of the parameter mismatch. This system
also is characterized by a significant number of boxes with a probability
between $0.4$ and $0.6$. 

Figure \ref{fig:S3_res} presents results for S3. The range of considered
excitation frequency and amplitude correspond to the values analyzed
in the original paper \cite{Wang2024}. We can see on all panels
that the region with the highest probability (boxes in shades of red)
is not as uniform as in the two previous cases. For S3 in the shade
red areas, probability varies between $0.9$ and $1$. For S1 and
S2 there were uniform regions with probability equal to $1.$ Interestingly,
at the same time, there are only a few boxes where the probability
is between $0.3$ and $0.8$ meaning that in practice we can assume
only two possible outcomes (Major probability of reaching cross-well
motion and major probability of reaching intra-well motion), except
boundaries between these two regions where are the boxes with the
shades of light blue and yellow. Even more interesting is the fact
that parameter mismatch is not affecting significantly the probability
of reaching cross-well motion. All four panels are qualitatively similar,
implying that the system is resistant to parameter fluctuations. 

Figure \ref{fig:S4_res} shows results for S4. The range of considered
excitation frequency and amplitude correspond to the values analyzed
in the original paper \cite{Costa2024}. We observe that the cross-well
motion is obtained in a much greater range of frequencies than in
all previous examples. There are two tongue-shaped regions, the first
one on the left side of the plot ensures that the desired behavior
of the system is almost solely possible (with a probability $0.9$
or more). The second tongue-shaped region is more interwoven by different
attractors and the probability of reaching cross-well motion varies
between $0.3$ and $1$. Similarly as in the previous case (S3), this
system is resistant to parameter mismatch, meaning that the system
dynamics is not affected even if parameter values are misjudged by
$\pm10\%$. Between two tongue-shaped red shade regions there are
areas with drop of the probability up to $0.5$. From the practical
point of view, only the left red-shaded region provides stability
of the operation of the device in the desired manner. All other areas
on the plots suggest that the stability of the cross-well motion is
minor and it may easily jump to another type of stable motion. 

Our analysis revealed variations in how the four energy harvesters
responded to parameter mismatch. Electromagnetic harvesters (S1 and
S2) exhibited similar behavior, with S2 (being more complex) demonstrating
greater basin stability across a wider range. However, S2 also had
larger regions with a $50\%$ chance of cross-well motion, indicating
a higher probability of undesirable transitions despite overall basin
stability. Both S1 and S2 showed minimal impact from \textpm 2.5\%
parameter mismatch, but \textpm 10\% variations significantly reduced
the probability of reaching cross-well motion (up to 50\% decrease).
Piezoelectric harvesters (S3 and S4) displayed superior resistance
to parameter mismatch, maintaining stability in regions with dominant
basin attraction. This is evident in Figures \ref{fig:S3_res}, \ref{fig:S4_res},
and the darker blue shades in Figure \ref{fig:compare_prob} (panels
c \& d), indicating maximum absolute stability changes of 23\% (S3)
and 30\% (S4) even with \textpm 10\% mismatch. Maximum absolute differences
in basin stability for different systems are gathered in the table
below:
\begin{center}
\begin{tabular}{c|>{\centering}p{2cm}|>{\centering}p{2cm}|>{\centering}p{2cm}|>{\centering}p{2cm}}
\multirow{2}{*}{Degree of parameter mismatch} & \multicolumn{4}{c}{Maximum absolute basin stability change}\tabularnewline
\cline{2-5} \cline{3-5} \cline{4-5} \cline{5-5} 
 & S1 & S2 & S3 & S4\tabularnewline
\hline 
$\pm2.5\%$ & $0.75$ & $0.23$ & 0.16 & $0.22$\tabularnewline
\hline 
$\pm5\%$ & $0.74$ & $0.30$ & $0.27$ & $0.23$\tabularnewline
\hline 
$\pm10\%$ & 0.75 & $0.54$ & 0.23 & 0.30\tabularnewline
\hline 
\end{tabular}
\par\end{center}

\section{Conclusions}

This study investigated the influence of parameter mismatch on the
stability of cross-well motion in four energy harvesters using a basin
stability metric. We showed how different percentages of mismatch
alter the system dynamics. Our results demonstrate that piezoelectric
harvesters exhibited greater resilience to parameter variations compared
to electromagnetic designs, which should be further investigated.
Safe operating ranges, in considered parameter space, identified through
basin stability analysis offer valuable insights for engineers and
scientists to ensure stable cross-well motion despite potential parameter
uncertainties. These findings highlight the influence of parameter
mismatch on the design and operation of energy harvesters.

\section*{Acknowledgment}

This paper has been completed while the first author was the Doctoral
Candidate in the Interdisciplinary Doctoral School at the Lodz University
of Technology, Poland.

This work is funded by the National Science Center Poland based on
the decision number 2018/31/D/ST8/02439.


\end{document}